\documentclass[aps,floatfix,twocolumn,showpacs,preprintnumbers]{revtex4}
\usepackage{graphicx}
\usepackage{bm}
\usepackage{amsmath,amssymb}
\usepackage{xcolor}

\graphicspath{ {./}{figs/}}
\begin{document}

\title{Calculated Unconventional Superconductivity via Charge Fluctuations
in Kagome Metal CsV$_{3}$Sb$_{5}$}
\author{Yuan Tian, Sergey Y. Savrasov}
\affiliation{Department of Physics, University of California, Davis, CA 95616, USA}

\begin{abstract}
Electrons on Kagome lattice exhibit a wealth of features including Dirac
points, van Hove singularities and flatbands. When the Fermi level is placed
at the van Hove saddle point, the Fermi surface is perfectly nested and a
rich variety of electronic instabilities is known to occur. The material
realization of such scenario is a recently discovered Kagome system CsV$_{3}$%
Sb$_{5}$ whose superconductivity near charge--density wave instability at
low temperatures points to an unconventional, non--electron--phonon, pairing
mechanism. Here we use a recently developed combination of density
functional theory with momentum and frequency--resolved self--energies
deduced from the so--called fluctuational--exchange--type random phase
approximation to study charge fluctuation mediated pairing tendencies in CsV$%
_{3}$Sb$_{5}$. Based on our numerical diagonalization of the BCS gap
equation, two competing solutions emerge from these calculations with $%
A_{1g} $ (anisotropic s-wave--like) and $B_{2g}$ ($d_{x^{2}-y^{2}}$,$d_{xy}$%
--like) symmetries of the superconducting order parameter. Our evaluated
Eliashberg spectral functions $\alpha ^{2}F(\omega )$ are purely due to
electronic correlations; they were found to be strongly peaked in the
vicinity of frequency 7 meV that sets the scale of charge fluctuations. The
superconducting coupling constants for the leading pairing channels are
estimated as a function the nearest neighbor Coulomb interaction $V,$ a
well--known prime parameter of the extended Hubbard model. They were found
in the range of 0.2-0.4 depending on $V$. We evaluate the superconducting\ $%
T_{c}$ close to the values that are observed experimentally that point to
the charge fluctuations to provide a substantial contribution to the pairing
mechanism in CsV$_{3}$Sb$_{5}.$
\end{abstract}

\maketitle

Unconventional mechanisms of superconductivity have always been a subject of
intense interest in the research of quantum materials with such much
celebrated examples as high--temperature superconducting cuprates\cite%
{Cuprates}, ironates\cite{Ironates} and recently, nickelates\cite{Nickelates}%
, where antiferromagnetic spin fluctuations are thought to be the primary
source of the Cooper pairing\cite{HTC-1,HTC-2}. There is however another
class of systems, for which the proximity not to the spin but to the charge
density wave (CDW) instability can be linked to the formation of the Cooper
pairs. The most notable example is the 30K superconductivity in potassium
doped BaBiO$_{3}$\cite{BKBO} where the Bi ions in the parent insulating
compound exist in a charge disproportionated state.

The discovery\cite{CsV3Sb5} of a family of non--magnetic metals CsV$_{3}$Sb$%
_{5}$, KV$_{3}$Sb$_{5}$, RbV$_{3}$Sb$_{5}$ with vanadium ions forming a
Kagome lattice framework is currently generating a great interest due to the
appearance of multiple charge--ordered states at high temperatures\cite%
{ChargeOrders}, as well as of bulk superconductivity with T$_{c}$=2.5K in CsV%
$_{3}$Sb$_{5}$\cite{CsV3Sb5Supra} and with T$_{c}$=0.8K in KV$_{3}$Sb$_{5}$%
\cite{KV3Sb5Supra} whose normal state electronic structure categorizes these
systems as Z$_{2}$ topological metals\cite{CsV3Sb5Supra}. Remarkably,
pressure dependent studies of CsV$_{3}$Sb$_{5}$ have shown that the CDW
order can be suppressed by applying a 2GPa pressure which leads to the
enhanced T$_{c}$ value of 8K\cite{CsV3Sb5Pressure}.

Although the observed small values of $T_{c}$ are well within the reach of
the conventional mechanism, theoretically calculated electron--phonon
coupling constants are generally found to be small\cite{EPI-PRL}. There
exists a factor--of--two discrepancy between predictions of the theory and
the kinks in the band dispersions of the Fermi electrons measured by Angle
Resolved Photoemission Spectroscopy (ARPES)\cite{EPI-ARPES}. More
importantly, several experiments performed on CsV$_{3}$Sb$_{5}$ point to a
strong momentum dependence of the superconducting energy gap. Scanning
tunneling microscopy (STM)\cite{STM-PRX,STM-PRL,STM-Nature} has detected a
V--shaped density of superconducting states, indicating the presence of
nodes in the order parameter. On the other hard, the magnetic penetration
depth experiments\cite{SW-NPJQM,SW-PRR,SW-NatureComm, SW-Nano} suggest
nodeless but anisotropic superconductivity. Despite been presently
controversial, both observations are incompatible with the constant energy
gaps which are observed in the vast majority of electron--phonon
superconductors and indicate the unconventional nature of the pairing state.

\begin{figure*}[tbp]
\includegraphics[height=0.413\textwidth,width=1.0\textwidth]{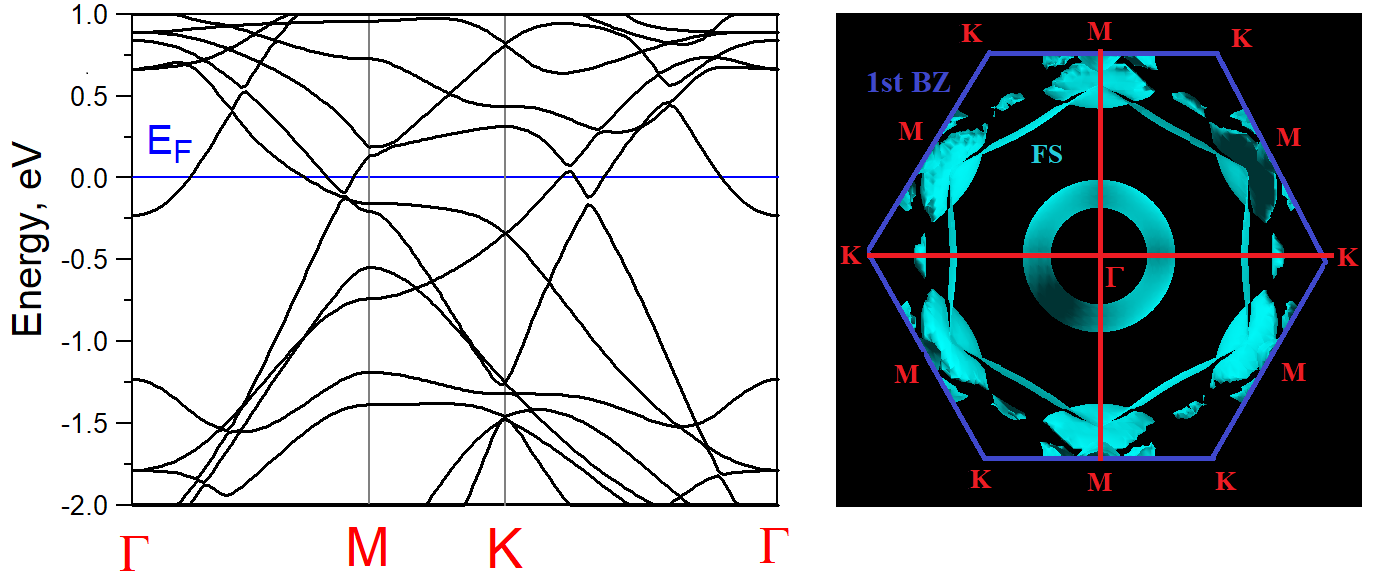}
\caption{Density functional calculations for CsV$_{3}$Sb$_{5}$: (a) the
electronic energy bands with clearly distinguished VHS saddle point below
the Fermi level at $M,$ the Dirac point at $K,$ and nearly dispersionless
bands at 1 eV above the $E_{F}$. b) The Fermi surface with the visible
hexagonal pattern characteristic of the tight--binding model on Kagome
lattice known for its nesting features along $\Gamma M.$ }
\label{FigBands}
\end{figure*}

Electronic instabilities on the Kagome lattice have been a subject of recent
theoretical works. Its minimal three--band model of itinerant electrons with
short--range hoppings is known for a wealth of features including the
existence of a Dirac point at the Brillouin Zone (BZ) point $K$=(1/3,1/$%
\sqrt{3}$,0)2$\pi /a$ of hexagonal lattice, a van Hove singularity (VHS) at
the point $M$=(0,1/$\sqrt{3}$,0)2$\pi /a$, and$\ $a dispersionless (flat)
band for all wavevectors. When the Fermi level is pinned at the VHS, the
Fermi surface is represented by a perfect hexagon and becomes nested along
the whole line $\Gamma M$ of the BZ. In this regime, the extended Hubbard
model with the on--site and intersite Coulomb interactions $U$ and $V$ has
been studied using a variational cluster approach\cite{JXLi} and a
functional renormalization group theory\cite{QHWang,Thomale}; it reveals a
rich variety of quantum phases including the appearance of magnetic order,
charge density waves and superconductivity. The symmetries of the
superconducting order parameter recovered from these studies include
anisotropic s--wave state and $d_{x^{2}-y^{2}},d_{xy}$ two--fold degenerate
state, for which a fully gapped chiral combination $d_{x^{2}-y^{2}}+id_{xy}$
was found to be energetically most favorable. Later studies of the Kagome
based tight--binding models included renormalization group supplemented with
Landau theory analysis of various charge density wave modulations seen in CsV%
$_{3}$Sb$_{5}$ \cite{Balents}, and the low--energy effective model for
various CDW\ induced flux phases \cite{JHu} to explain the observed
time--reversal symmetry breaking in KV$_{3}$Sb$_{5}$\cite{TRSHasan}.

To address the issue of unconventional pairing state in CsV$_{3}$Sb$_{5}$,
here we use our recently developed approach\cite{LDA+FLEX,Hg-FLEX} that
evaluates superconducting pairing functions directly from first--principle
electronic structure calculations of the studied material using realistic
energy bands and wave functions available from density functional theory
(DFT)\cite{DFT}. The electronic self--energies are calculated for a manifold
of correlated electrons similar to as it is done in popular DFT+DMFT approach%
\cite{DFT+DMFT}, but they acquire full momentum and frequency resolution
within this method, which diagrammatically corresponds to the so--called
fluctuational--exchange (FLEX) type \cite{FLEX} random phase approximation
(RPA) incorporating all type of nesting--driven instabilities in the charge
and spin susceptibilities. The description of unconventional superconductors
in a realistic material framework without reliance on the tight--binding
approximations of the electronic structures became possible using this
DFT+FLEX(RPA) method. Our most recent applications\cite{Hg-FLEX} to HgBa$%
_{2} $CuO$_{4}$, a prototype single--layer cuprate superconductor, easily
recovered a much celebrated $d_{x^{2}-y^{2}}$ symmetry of the order
parameter, and the prediction of competing $s_{\pm },d_{xy}$ pairing states
was given\cite{La-FLEX} for a recently discovered high--temperature
superconducting nickelate compound La$_{3}$Ni$_{2}$O$_{7}$.

The main physical picture emergent from the experimental data for CsV$_{3}$Sb%
$_{5},$ is that a nearly singular behavior in the charge susceptibility
plays an important role in the $2.5K$ superconductivity at ambient pressure%
\cite{CsV3Sb5Supra}, where a variety of charge ordered phases is seen at
higher temperatues\cite{STM-Nature}, and that the rise of $T_{c}$ to 8K by
applying a pressure of 2GPa \cite{CsV3Sb5Pressure} is related to the
suppression of the CDW. This prompts us to consider the on--site and the
neighboring--site Coulomb interaction parameters $U$ and $V$ for vanadium $d$%
--electrons to be of the same order of magnitude with the parameter $V$
tuning the system to the instability point to allow strong charge
fluctuations.

\begin{figure*}[tbp]
\includegraphics[height=0.317\textwidth,width=1.0\textwidth]{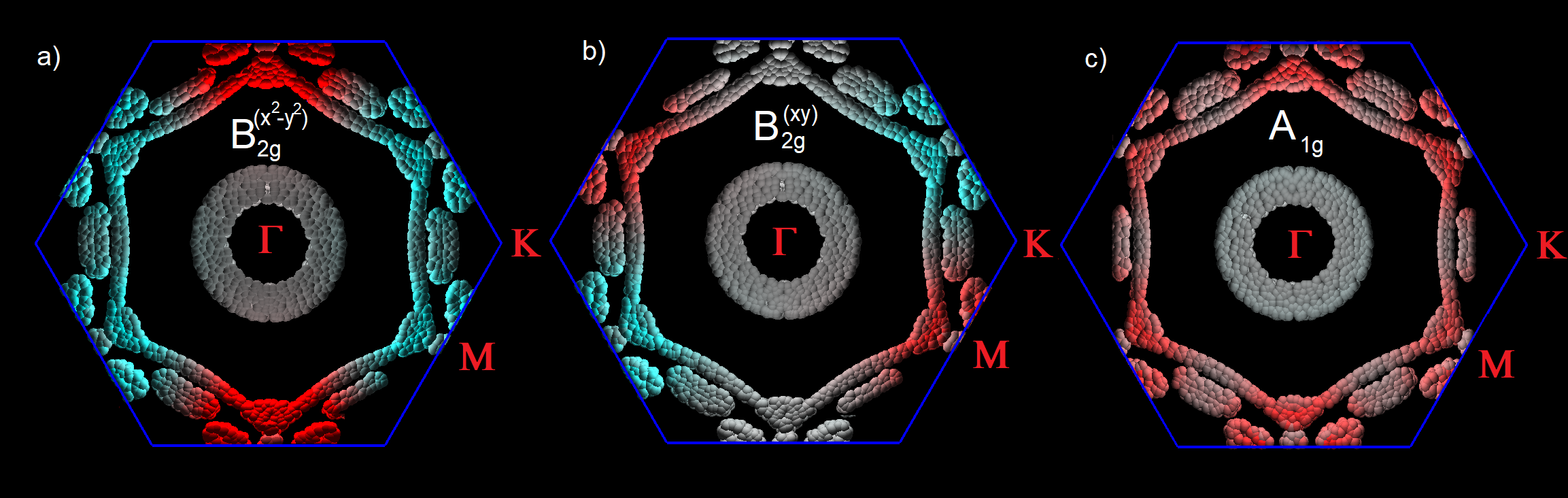}
\caption{{}Calculated superconducting energy gap $\Delta (\mathbf{k}j)$ for
singlet pairing in CsV$_{3}$Sb$_{5}$ using numerical solution of the
linearized BCS gap equation with the pairing interaction evaluated using the
DFT+FLEX(RPA)\ approach. Blue/red color corresponds to the negative/positive
values of $\Delta (\mathbf{k}j).$ Plots a) and b) correspond to $B_{2g}$
symmetry ($d_{x^{2}-y^{2}}$ and $d_{xy}$ like); plot c) corresponds to $%
A_{1g}$ symmetry (anisotropic s--wave--like).}
\label{FigGaps}
\end{figure*}

To uncover whether such charge fluctuational mechanism\ can explain or
contribute to superconductivity in CsV$_{3}$Sb$_{5}$, we numerically
evaluate the pairing interaction describing the scattering of the Cooper
pairs as a function of the intersite $V$ while fixing the on--site $U\ $to
its representative value of 0.1 Ry (=1.36 eV). This pairing function is then
used to exactly diagonalize the linearized Bardeen--Cooper--Schrieffer (BCS)
gap equation on a three--dimensional k--grid of the Fermi points in the BZ.
The highest eigenvalue $\lambda _{\max }$ deduced from this procedure
represents a coupling constant similar to the electron--phonon $\lambda $ in
conventional theory of superconductivity. We generally find $\lambda _{\max }
$ to be negligible unless $V$ is tuned to the close proximity to the CDW
occurring in our procedure at around 1.8 eV. We recover two nearly
degenerate solutions of the superconducting order parameter\ from these
calculations: first, of $B_{2g}$ ($d_{x^{2}-y^{2}},d_{xy}$--like) and,
second, of $A_{1g}$ (anisotropic s-wave--like) symmetry. Using spectral
representation for the pairing interaction, we evaluate charge fluctuation
induced Eliashberg spectral functions $\alpha ^{2}F(\omega )$ which were
found to be strongly peaked at the frequency 7 meV. To allow estimates for
the $T_{c}$, charge fluctuational contribution to the electronic mass
enhancement $m^{\ast }/m_{band}=1+\lambda _{cf}$ $\ $is evaluated$\ $to
produce the effective coupling constants $\lambda _{eff}=\lambda _{\max
}/(1+\lambda _{cf})$ for the leading paring channels. These were found in
the range of 0.2--0.4 depending on $V$ and leads to the\ $T_{c}\ $estimates
close to the values that are observed experimentally.

We perform our density--functional electronic--structure calculations using
the full potential linear muffin--tin orbital method\cite{FPLMTO}. The
result for the electronic energy bands is shown in Fig. \ref{FigBands}(a)
along major high--symmetry directions of the hexagonal BZ. In accord with
the previous study \cite{CsV3Sb5Supra}, it shows the band dispersions in the
vicinity of the Fermi level $E_{F}$ originating from the vanadium
d--orbitals. Despite their complexity, the VHS saddle point just below the
Fermi level at point $M,$ the Dirac point at $K,$ and the nearly
dispersionless bands at 1 eV above the $E_{F}$ are clearly distinguished.
The Fermi surface shown in Fig. \ref{FigBands}(b) is quasi--two dimensional
with the visible hexagonal pattern. All these features are characteristic of
the minimal tight--binding model on Kagome lattice known for its nesting
along $\Gamma M.$

We further utilize our DFT+ FLEX(RPA) method to evaluate the charge
fluctuation mediated pairing interaction. The Fermi surface is
triangularized onto small areas described by about 6,000 Fermi surface
momenta for which the matrix elements of scattering between the Cooper pairs
are calculated using the approach described in Ref. \cite{Hg-FLEX}. The
linearized BCS gap equation is then exactly diagonalized and the set of
eigenstates is obtained for both singlet ($S=0$) and triplet ($S=1$) Cooper
pairs. The highest eigenvalue $\lambda _{\max }$ represents the physical
solution and the eigenvector corresponds to superconducting energy gap $%
\Delta (\mathbf{k}j)$ where $\mathbf{k}$ is the Fermi surface momentum and $%
j $ numerates the Fermi surface sheets.

We find that there are three highest eigenvalues that appear very close to
each other. The leading pairing channel is two--fold degenerate and the
subleading one is non--degenerate with its eigenvalue appearing only within
6\% of the maximum eigenvalue. We analyze the behavior of $\Delta (\mathbf{k}%
j)$ as a function of the Fermi momentum using the values of $U$=1.36 eV and $%
V$ = 1.75 eV. The solutions are related to the spin singlet states, and Fig. %
\ref{FigGaps}(a),(b) shows the behavior of the two--fold degenerate $\Delta (%
\mathbf{k}j)$, while Fig. \ref{FigGaps}(c) corresponds to the
non--degenerate one. One can see that the two--fold degenerate eigenstate
shows the behavior corresponding to the $B_{2g}$ symmetry ($d_{x^{2}-y^{2}}$%
--like in a) and $d_{xy}$--like in b), The plot distinguishes negative and
positive values of $\Delta (\mathbf{k}j)$ by blue and red colors while zeros
of the gap function are colored in grey. The non--degenerate solution of $%
A_{1g}$ symmetry is plotted in Fig. \ref{FigGaps}(c) where the gap function
is strongly anisotropic exhibiting its maxima close to the M points of the
BZ and nearly zeroes in between.

\begin{figure}[tbp]
\includegraphics[height=0.388\textwidth,width=0.40%
\textwidth]{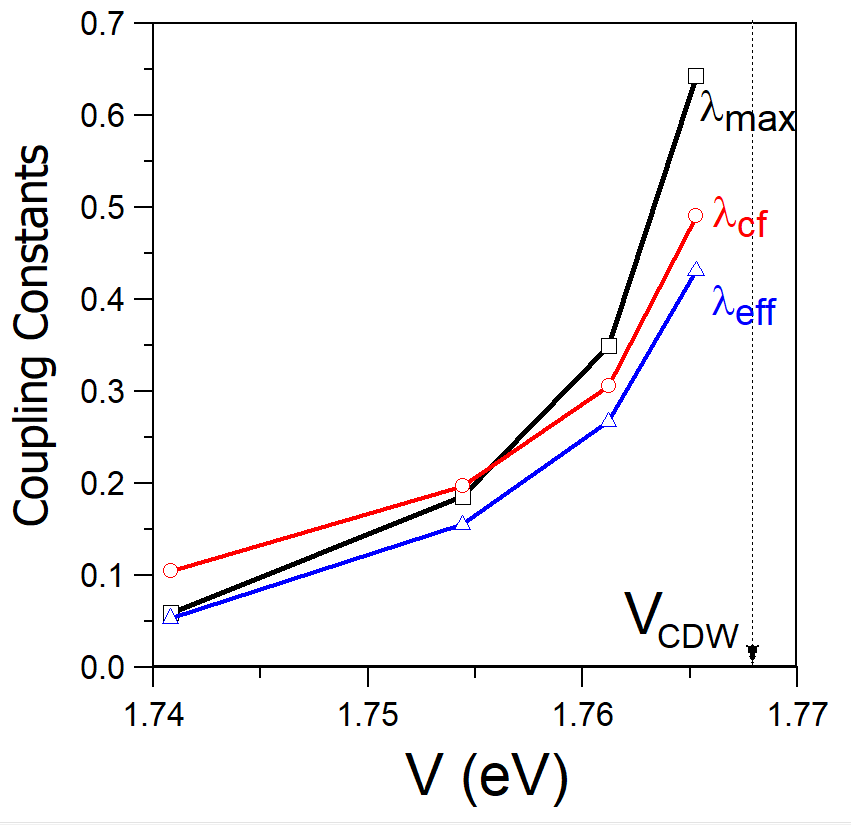}
\caption{{}Calculated using DFT+FLEX(RPA)\ method dependence of the highest
eigenvalue $\protect\lambda _{\max }$ (squares connected by black lines)
corresponding to $B_{2g}$ symmettry of the linearized BCS equation, as well
as charge fluctuational mass enhancement parameter $\protect\lambda _{cf}$
(cirles connected by red lines) as a function of the inter--site Hubbard
interaction $V$ close to its critical value $V_{CDW}=$ 1.768eV for the
vanadium d-electrons in CsV$_{3}$Sb$_{5}$. The effective coupling constant $%
\protect\lambda _{eff}=\protect\lambda _{\max }/(1+\protect\lambda _{cf})$
determining the strength of the charge fluctuational pairing is also shown
(triangles connected by blue lines).}
\label{FigLambdas}
\end{figure}
We can gain additional insight on the behavior of the eigenvalues by varying
the intersite Coulomb interaction $V.$ We first evaluate the divergency of
the charge susceptibility that occurs at $V_{CDW}$ slightly less than 1.77 eV%
$.$ We use the range of values $V<V_{CDW}$ to extract from the BCS\ gap
equation the behavior of the highest eigenstates $\lambda _{\max }$ and
their symmetries as a function of $V$. \ We find that both $B_{2g}$ and $%
A_{1g}$ symmetries robustly dominate over all other solutions with the
two--fold degenerate $B_{2g}$ state to be only 6\% larger than the
non--degenerate one. Obviously, however, if the electron--phonon coupling
constant $\lambda _{e-p}\ \approx 0.25$\cite{EPI-PRL} is taken into account,
the $A_{1g}$ pairing will become the leading one.

It is interesting to discuss the dependence of $\lambda _{\max }$ as a
function of $V$ that is shown in Fig. \ref{FigLambdas} (squares connected by
black lines). We see that the values of $\lambda _{\max }$ are very small
unless $V$ approaches closely to $V_{CDW}$ where it reaches the values
0.2--0.6. This indicates that the charge fluctuations produce essential
contribution to the pairing only in the immediate vicinity of the CDW
instability.

To get insight on possible range of critical temperatures that can be
obtained using the charge fluctuation mechanism, we recall that it is not
the eigenvalue $\lambda _{\max }$ but the effective coupling constant $%
\lambda _{eff}$ enters the BCS $T_{c}$ expression: $T_{c}\approx \omega
_{cut}\exp (-1/\lambda _{eff}).$ If we neglect phonons, the cutoff frequency 
$\omega _{cut}$ is thought here due to charge fluctuations, and $\lambda
_{eff}$ incorporates the effects associated with the mass renormalization
describing by the parameter $\lambda _{cf}.$ It should also be weakened
somewhat by the Coulomb pseudopotential $\mu _{m}^{\ast }$ which should
refer to the same pairing symmetry $m$ as $\lambda _{\max }$: 
\begin{equation}
\lambda _{eff}=\frac{\lambda _{\max }-\mu _{m}^{\ast }}{1+\lambda _{cf}}
\label{Leff}
\end{equation}

The mass enhancement can be evaluated as the Fermi surface (FS) average of
the electronic self--energy derivative taken at the Fermi level%
\begin{equation}
\lambda _{cf}=-\langle \frac{\partial \Sigma (\mathbf{k},\omega )}{\partial
\omega }|_{\omega =0}\rangle _{FS}  \label{Lcf}
\end{equation}%
Our calculated dependence of $\lambda _{cf}\ $on $V$ is shown in Fig. \ref%
{FigLambdas} (circles connected by red lines). It is seen to exhibit the
behavior very similar to $\lambda _{\max }:$ the values of $\lambda _{cf}$
are found to be modest unless $V$ lies in the vicinity of $V_{CDW}$ where $%
\lambda _{cf}$ is found between 0.2 and 0.5.

To give estimates for the effective coupling constant, $\lambda _{eff},$ we
notice that $\mu _{m}^{\ast }$ is expected to be very small for the pairing
symmetries different from the standard s--wave \cite{Alexandrov}. We
therefore expect this parameter to be zero for the $B_{2g}$ pairing state.
The plot of $\lambda _{eff}=\lambda _{\max }/(1+\lambda _{cf})$ vs. $V$\ is
shown in Fig \ref{FigLambdas}. One can see that the range of these values is
between 0.2 and 0.4 in the proximity to the CDW.

One can easily incorporate the electron--phonon $\lambda _{e-p}\approx 0.25$%
\cite{EPI-PRL} into this discussion. The kink in the band dispersion of the
Fermi electrons will become a sum $\lambda _{cf}+\lambda _{e-p}.$ It
essentially doubles as compared to the individual contributions due to
phonons or charge fluctuations and will match the recent ARPES study of the
electronic mass enhancement in CsV$_{3}$Sb$_{5}$\cite{EPI-ARPES}. Our
estimate for the $\lambda _{eff}$\ in the $B_{2g}$ pairing state will be
lower by 15\% or so due to a slightly larger denominator in Eq.(\ref{Leff}).
For the $A_{1g}$ pairing state, our estimated $\lambda _{\max }$ due to
charge fluctuations should be supplemented with $\lambda _{e-p}\approx 0.25\ 
$but the value of $\mu ^{\ast }\approx 0.12$ should be taken into account in
Eq.(\ref{Leff}). This will increase our estimate for $\lambda _{eff}$\ by
10\%t or so.

\begin{figure}[tbp]
\includegraphics[height=0.315\textwidth,width=0.40\textwidth]{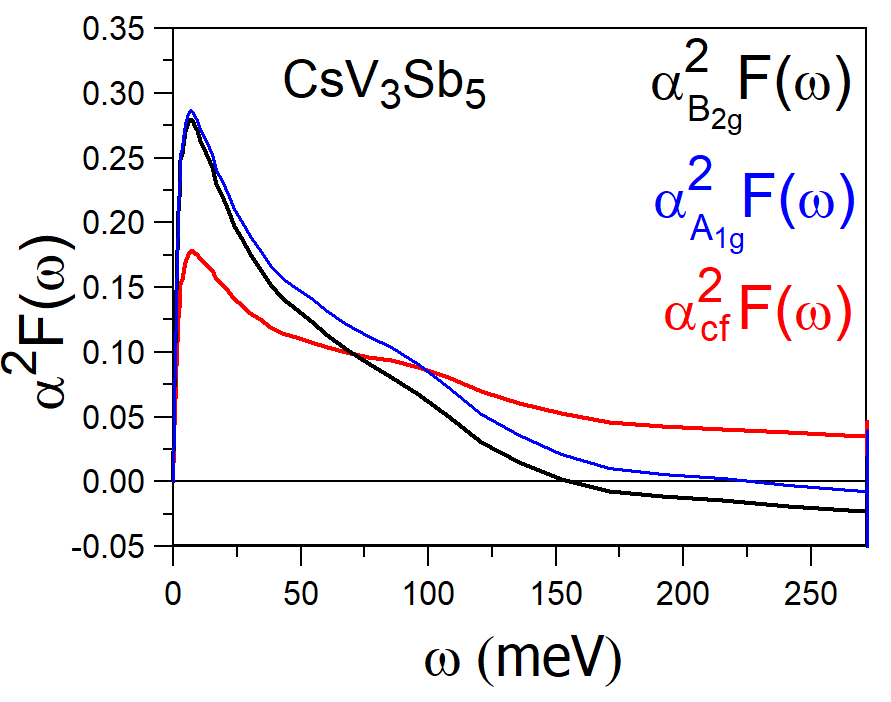}
\caption{Calculated Eliashberg spectral functions $\protect\alpha ^{2}F(%
\protect\omega )$ due to charge fluctuations for the leading pairing
symmetries, $B_{2g}$ (black lines), and $A_{1g}$ (blue lines) of the
superconducting order parameter in CsV$_{3}$Sb$_{5}.$The spectral function $%
\protect\alpha _{cf}^{2}F(\protect\omega )$ whose inverse frequency moment
produces charge fluctuational mass enhancement parameter $\protect\lambda %
_{cf}$ is shown in red.}
\label{FigA2F}
\end{figure}
To obtain the estimates for the range of charge fluctuational energies that
set the scale $\omega _{cut}\ $in the BCS T$_{c}$ expression, we calculate
Eliashberg spectral functions $\alpha ^{2}F(\omega )$ that are responsible
for the pairing. Since in our method the Cooper pairs scatter on the
statically screened Coulomb interaction, Re$K_{\mathbf{k}j\mathbf{,k}%
^{\prime }j^{\prime }}(0),$ the Kramers--Kroenig transformation provides its
frequency resolution, proportional to Im$K_{\mathbf{k}j\mathbf{,k}^{\prime
}j^{\prime }}(\omega )/\omega .$ The frequency resolved function Im$K_{%
\mathbf{k}j\mathbf{,k}^{\prime }j^{\prime }}(\omega )$ is averaged over the
eigenvectors of the BCS\ gap equation $\Delta (\mathbf{k}j),\Delta (\mathbf{k%
}^{\prime }j^{\prime })$ to give rise to the superconducting $\alpha
^{2}F(\omega ),$ whose double inverse frequency moment evaluates the
eigenvalue $\lambda _{\max }$. We perform this procedure for the leading
eigenstates of $B_{2g}$ and $A_{1g}$ symmetries, and also calculate the
spectral function $\alpha _{cf}^{2}F(\omega )$ whose inverse moment produces
the mass enhancement parameter $\lambda _{cf}.$

We present this data in Fig.\ref{FigA2F}, where the black/blue lines show
the superconducting $\alpha ^{2}F(\omega )$ of the $B_{2g}$ and $A_{1g}$
symmetries, respectively, while the red line describes the charge
fluctuational $\alpha _{cf}^{2}F(\omega ).$ A strong peak at the frequencies
around 7 meV is seen in all three plots and indicates a characteristic
frequency of the charge fluctuations. This can be compared to the Debye
frequency of 142K=12 meV deduced from the calculated phonon spectrum of CsV$%
_{3}$Sb$_{5}$\cite{EPI-PRL}.

We can judge about the values of $T_{c}$ obtained within the charge
fluctuational mechanism using our estimated $\omega _{cut}\approx $ 7 meV
and the values of $\lambda _{eff}$ that we calculate in Fig. \ref{FigLambdas}%
. For $\lambda _{eff}=0.2,$ the BCS$\ T_{c}\approx \omega _{cut}\exp
(-1/\lambda _{eff})\approx $ 0.5K. Once we get closer to the CDW
instability, the effective coupling increases to the values 0.4 and the
corresponding BCS $T_{c}\approx $ 7K. Should the contribution from the
phonons be included in the $A_{1g}$ pairing channel, a refined estimate for $%
\lambda _{eff}$ is inflated by about 10\% and the BCS $T_{c}^{\prime }s$ are
about 20\% larger than the values quoted above. Given the exponential
sensitivity$,$ these estimates are clearly within the range of the $T_{c}$'s
observed experimentally.

In conclusion, we numerically estimated the charge fluctuation mediated
pairing tendencies in the recently discovered Kagome metal CsV$_{3}$Sb$_{5}.$
These cacluations are done directly using first--principle electronic
structures without resorting to tight--binding approximations of any kind.
Two competing pairing channels have been recovered in our study: the
two--fold degenerate $B_{2g}$ nodal states ($d_{x^{2}-y^{2}},d_{xy}$ --like)
and the non--degenerate $A_{1g}$ nodeless state (anisotropic s--wave--like)
with the effective coupling constants in the range 0.2--0.4 depending on the
strength of the nearest neighbor Coulomb repulsion between the vanadium $d$%
--electrons. Similar values are predicted to contribute to the electronic
mass enhancement due to charge fluctuations. These estimates provide
substantial contributions both to the electronic kinks of the Fermi
electrons in the normal state, and to the strength of the Cooper pairing in
the superconducting state.

\end{document}